\documentclass[intlimits,twoside,a4paper]{article}

\usepackage[eqsecnum]{cmpj3}
\usepackage[utf8]{inputenc}

\issue{2026}{29}{1}{13803}
\doinumber{10.5488/CMP.29.13803}
\title[Vortex formation]
{Vortex formation in the Vicsek model with internal chirality of self-propelling objects
}
\author[W. T. Gozdz, A. Ciach]{W. T. Gozdz\orcid{0000-0003-4506-6831},
       A. Ciach\orcid{0000-0002-5556-401X}\thanks{Corresponding author: \email{aciach@ichf.edu.pl}.}}
\address{
Institute of Physical Chemistry,  Polish Academy of Sciences, 01-224 Warszawa, Poland
}
\date{Received 11 December 2025; revised 17 January 2026; accepted 19 January 2026; published 30 March 2026}

\begin{document}

\maketitle

\begin{abstract}
Effect of internal chirality on collective motion of  a   
large number of active objects is studied by simulations of 
appropriately modified Vicsek model. We add a fixed angle to 
the noise and consider small ratios, $p$,
 between this angle and the maximal deviation 
from the average local direction of motion.
 When the above ratio is $p=1/120$, the 
traveling bands observed with the symmetrical noise are
 destroyed, and small bands moving in different directions
 appear. Circular rotating flocks of objects with the same 
orientation are formed for $p=1/7.5$. Stable 
vortexes in the stationary state were found from $p=1/60$ to $p=1/20$. Velocity autocorrelation function shows 
equilibrium between the inflow and the outflow to and from 
the vortex. Long-time evolution is considerably influenced 
by a temporary trapping of the objects in the vortex. 
The ballistic behavior for the symmetrical noise changes to the diffusive
 behavior for the chirality leading to the onset of vortexes.
\end{abstract}

 \section{Introduction}
\label{intro}

In biological systems, many active microorganisms like bacteria, algae, active cells such as spermatozoa, lymphocytes, or organisms like insects, birds, fish, and herds of other animals, as well as artificial microswimmers can exhibit complex motion~\cite{toner:98:0,bechinger:16:0,mora:16:0,peruani:12:0,afroze:21:0,
	reza:20:0,velho:21:0,cholakova:21:0,tan:22:0,kreienkamp:22:0,breier:16:0,
	hoffmann:25:0}. Complex spatiotemporal organization in ensembles of self-propelled objects includes flocking, swirling, travelling bands formation and phase separation~\cite{toner:98:0,bechinger:16:0,mora:16:0,peruani:12:0,
afroze:21:0,reza:20:0}. The movement of individual objects in the ensemble can be governed by the interactions with the environment or with other members of the ensemble. 

 In precise theoretical models of active objects moving in a solvent or on interfaces, the structure and hydrodynamics of the solvent are considered in addition to the movement of the objects~\cite{kitahata:23:0,sprenger:20:0,reza:20:0}; thus, the numerical complexity of this type of modelling restricts the study to a small number of active particles. On the other hand, the collective motion of large numbers of active objects can have similar features regardless of the mechanism of propulsion and the nature of the objects, as observed in flocks of birds or schools of fish. Thus, the pattern formation by a large number of active objects can be studied by highly simplified models, such as the Vicsek model~\cite{vicsek1995novel} and its variations~\cite{chate2008collective,chate2008modeling,gregoire2004onset}. 
 
In this work we focus on the influence of internal chirality  of self-propelling objects on synchronization of the motion of a very large number of these objects moving on an interface or a surface. Here by the chirality we mean a preferential rotation of the self-propelled objects to the left or to the right side of the object.
Internal chirality can be induced for example by an asymmetric shape or by different lengths of two flagella in a bacterium, and several models of self-propelling objects with internal chirality were introduced and studied~\cite{afroze:21:0,zhang:22:0,sahala:25:0,liebchen:22:0}.  Notable examples of self-propelling chiral objects in nature include many  biological active objects, such as ensembles of microtubuli~\cite{Sumino2012}, E. coli 
	close to walls and interfaces~\cite{berg:90:0},  magnetotactic bacteria in rotating external
	fields~\cite{erglis:07:0}, and sperm cells~\cite{riedel:05:0}. Some synthetic swimmers, such as
	L-shaped self-phoretic swimmers~\cite{tenHagen:14:0} and actuated
	colloids  can be characterised by internal chirality as well.
	In a macroscale, vortex formation can be observed in schools of fish or in large swarms of insects.

To study the collective motion of very large groups of self-propelled chiral objects, we modify the Vicsek model and restrict ourselves to a two-dimensional (2D) system. 
We are particularly interested in the effect of chirality on the synchronized motion of the active objects in conditions leading to the formation of travelling bands in the original Vicsek model~\cite{serna:23:0}. We first study the spatio-temporal organization for different degrees of chirality, and then perform more detailed investigations for the range of chirality leading to the  formation of a vortex, or of several vortexes in very large systems. 

The model and the simulation method are described in section~\ref{model}. The results are presented in section~\ref{results}. In the final section~\ref{conclusions} we present concluding remarks.

 \section{Model and methods}\label{model}
In the 2D Vicsek model~\cite{vicsek1995novel}, the speed, $v_0$, of each self-propelled particle is constant, but the velocity vector depends on the direction of motion of close neighbors. More precisely, the collective motion follows from adjusting the velocity of each particle to the average velocity of the  particles whose distance from the considered particle is $r\leqslant r_c$.
 The radius $r_c$  and  $v_0$ define the unit length 
and speed, respectively. Time unit in the model is set to $\delta t  =r_c/v_0$. We use dimensionless length, speed and time, and assume $r_c = 1.0$, $v_0=1.0$ and $\delta t  = 1.0$.

The time evolution of the position of the $i$-th particle
is governed by the following equation,
\begin{equation}
 \boldsymbol{r}_i(t + \delta t) = \boldsymbol{r}_i (t) + \boldsymbol{v}_i(t)\delta t , 
 \label{e:VicsekX}
\end{equation}
where $\boldsymbol{r}_i(t)$ and $\boldsymbol{v}_i(t)$ denote the position and the velocity of the particle $i$ at time $t$. 
The angle formed by the velocity vector of the $i$-th particle with the $x$ axis is denoted by $\theta_i(t)$. The angle $\theta_i(t + \delta t)$ at the next time step is calculated according to the following formula 
\begin{equation}
 \theta_i(t + \delta t) = \langle \theta (t) \rangle _{r_c} + \Delta\theta ,
 \label{e:VicsekTheta}
 \end{equation}
where $\langle \theta(t) \rangle_{r_c}$ is 
the average of the angles $\theta_j(t)$ with $j$ refering to all the particles located inside the circle of radius $r_c$ and the center at 
$\boldsymbol{r}_i (t)$. In (\ref{e:VicsekTheta}),  $\Delta\theta$ represents the deviation from the average angle at each time step (or 'noise') and in the original Vicsek model is given by $\Delta\theta=\Delta\theta_V= (2\xi-1)\theta_{\rm max}$, where $\xi$ is a random number uniformly distributed in the interval $[0,1]$ and $\theta_{\rm max}$ is the noise amplitude.

In order to study the effect of internal chirality of the active objects on the collective motion, we modify the noise in (\ref{e:VicsekTheta}) and postulate that in our modification of the Vicsek model
\begin{equation}
\label{theta_chi}
\Delta\theta= (2\xi-1)\theta_{\rm max}+\theta_\chi,
\end{equation}
 where $\theta_\chi$ is the degree of chirality that takes the same constant value for all particles. Similar model 
 was investigated for a small system with the number of particles of the order of 1000 in \cite{zhang:22:0}. For such a small number of particles, however, even the travelling bands in the systems without chirality are not formed. 
 
 We fix the noise amplitude  $\theta_{\rm max}$,  the number of particles $N$ as well as the number density $\rho$. The edge length of the square  simulation box, $L=\sqrt{N/\rho}$, is determined by $N$ and $\rho$.  We assume periodic boundary conditions. The simulations runs are of the length of  $10^5$ time steps to reach a steady state and  $10^5$ time steps to take averages. 
In this article, we present the results obtained for  
$N=810000$, $\rho =0.50$ and $\theta_{\rm max} = 1.0\,  {\rm rad} =({180}/{\piup})^\circ$.  In the initial configuration, the particles occupy sites of a square lattice with $900 \times 900$ lattice sites.
  For this set of parameters in the original Vicsek model, i.e., with the symmetrical noise, the particles move collectively, forming traveling bands. We verified that $N=810000$ is sufficiently large  and sufficiently small for formation of a single vortex in a system with small chirality.
  We restrict ourselves to the chirality $0.5^\circ\leqslant\theta_\chi\leqslant 8^\circ$, i.e., to quite small deviations from the symmetrical noise, and perform a more detailed analysis for  $1 ^\circ\leqslant\theta_\chi\leqslant 3^\circ$ leading to the formtion of a vortex.

To characterize the dynamic behaviour of the system, we compute the mean square displacement~(MSD) and the velocity auto-correlation function~(VAF) according to the following formulas:
\begin{equation}
{\rm MSD}(t)= \left\langle\Delta r(t)^2\right\rangle = \frac{1}{N}\sum_{i=1}^{N}\left[\boldsymbol{r}_i(t) - \boldsymbol{r}_i(0)\right]^2
 \label{e:MSD},
 \end{equation}
where the averaging $\langle ...\rangle$ in the first equality means averaging over trajectories of $N=810000$ particles, and
\begin{equation}
 {\rm VAF}(t) = \frac{1}{N}\left\langle \sum_{i=1}^{N}\left[\boldsymbol{v}_i(0) \cdot \boldsymbol{v}_i(t)\right]\right\rangle
 \label{e:VAF}, 
 \end{equation}
where the averaging $\langle ...\rangle$ is over 20 runs, $\boldsymbol{v}_i(t)$ is the velocity of the $i$-th particle at time $t$ and ``$\cdot$'' denotes the scalar product.

\section{Results}
\label{results}

\begin{figure}[h]
\centering\includegraphics[scale=1]{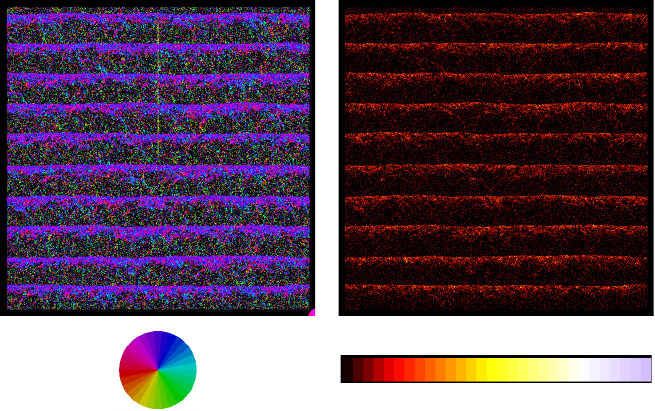}
\caption{(Colour online) Typical configuration in the stationary state  in the system with the density $\rho=0.5$, the number of particles  $N=810000$ and the chirality $\theta_\chi=0^\circ$. The velocity vectors of the particles (left-hand panel) are represented by small arrows with different color representing different  orientations of the particle. The color coding of the angle formed by the velocity vector with the horizontal axis  is presented in the circle below. 
The thin, vertical, yellow straight line in the center of the left panel shows the average direction of motion of all active objects. The local density of the active objects (right-hand panel) is represented by colorful spots, with the  density increasing from the dark to the bright color. The color coding is shown in the color map below the local densities. Note the elongated bands of objects moving in the same direction, and a small fraction of objects  dispersed outside the bands.}
\label{fig:bands} 
\end{figure}

 Our aim is to study the effect of the internal chirality on the  synchronised motion observed in the Vicsek model with symmetrical noise. Therefore, 
 we present the results obtained for  $N,\rho$ and $\theta_{\rm max}$ having the values for which self-assembly into traveling elongated bands takes place for $\theta_{\chi}=0^\circ$~\cite{serna:23:0,serna:25:0}  (see section~\ref{model}).  In figure~\ref{fig:bands},
a snapshot of a stationary state for $\theta_{\chi}=0^\circ$ is shown as a point of reference for the pattern formation with $\theta_{\chi}>0^\circ$.

\begin{figure}
\centering\includegraphics[scale=1]{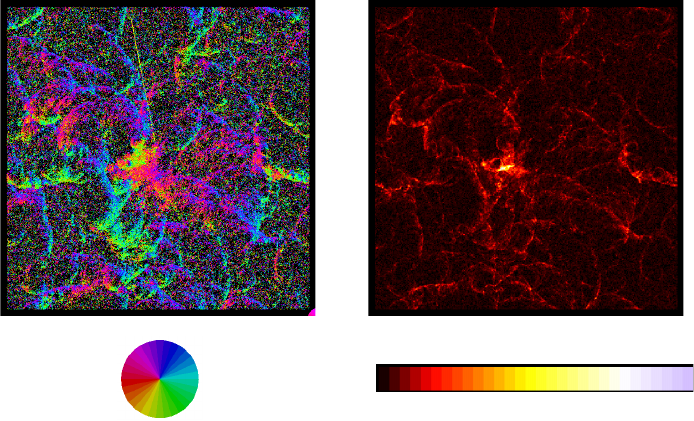}
\caption{(Colour online) Typical configuration in the stationary state  in the system with the density $\rho=0.5$, the number of particles  $N=810000$ and chirality $\theta_\chi=0.5^\circ$. In the left-hand panel, the velocity vectors of the particles are represented by small arrows with different color representing different  orientations of the particle. The color coding of the angle formed by the velocity vector with the horizontal axis  is presented in the circle below the shown velocities. In the right-hand panel, local density is shown for the same configuration, with the density coded according to the color map shown below the density distribution.  }
\label{fig:small}
\end{figure}

We have observed a strong dependence of the collective behavior of the active objects on the chirality~$\theta_\chi$. In the case of the smallest considered chirality, $\theta_\chi=0.5^\circ$,
 the  traveling bands that were observed  for $\theta_\chi=0^\circ$ in reference~\cite{serna:23:0} are broken, and as shown in figure~\ref{fig:small}, 
 isotropic structure consisting of small bands moving in different directions is obtained.
 
 \begin{figure}
 \centering\includegraphics[scale=1]{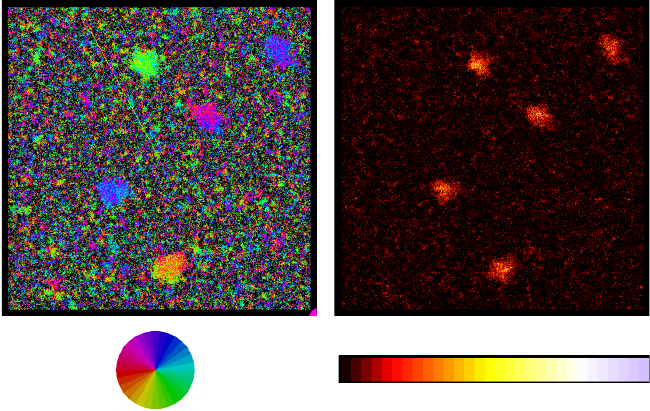}
\caption{(Colour online) The same as in figure~\ref{fig:small} but for  $\theta_\chi=8^\circ$. 
}
\label{fig:large}
\end{figure}
 
 Thus, already very small chirality has a disordering effect on the synchronized motion in one direction. In the opposite case of the largest considered chirality, $\theta_\chi=8^\circ$,  objects moving in the same direction form compact approximately circular flocks that rotate as a whole, as can be seen in figure~\ref{fig:large}.

\begin{figure}[h!]
  	\centering\includegraphics[scale=0.8]{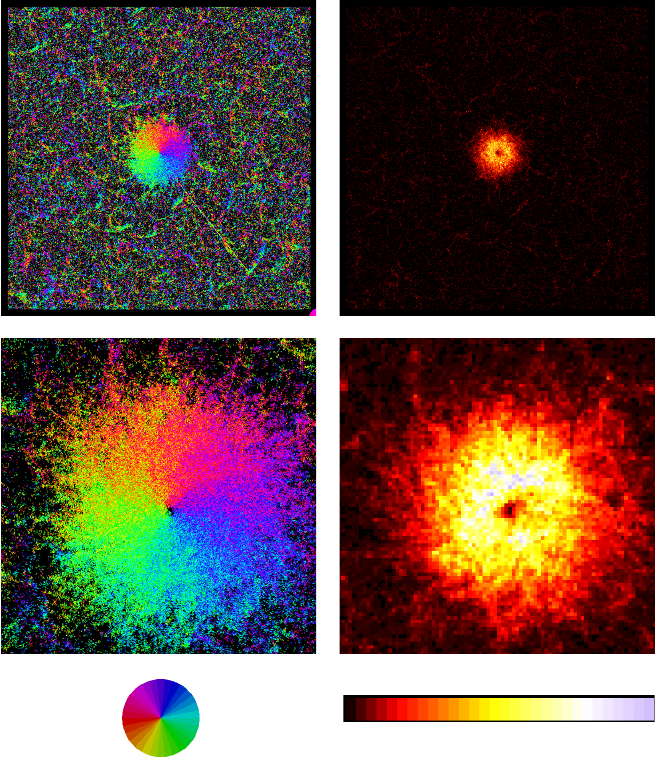}
  	\caption{(Colour online) Typical configuration in the stationary state  in the system with the density $\rho=0.5$, the number of particles  $N=810000$ and chirality $\theta_\chi=1^\circ$. 
  		Left-hand column:  velocities of the particles in the whole system (top)   and in a part of the system with the vortex zoomed in (bottom). The color coding of the velocity of the particles is shown in the circle below.
  		Right-hand column: local density in the same configuration in the whole system (top) and  in a part of the system with the vortex zoomed in (bottom).
  		The color coding of the local density is shown in the color map below. 
  	}
  	\label{fig:inter1}
  \end{figure}

\begin{figure}[h]
  	\centering\includegraphics[scale=1]{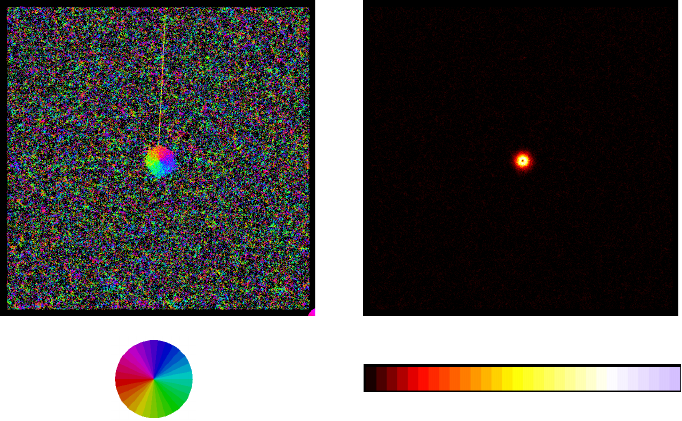}
  	\caption{(Colour online) Typical configuration  in the stationary state  in the system with the density $\rho=0.5$, the number of particles  $N=810000$ and chirality   $\theta_\chi=3^\circ$.  Velocities of the particles and local density are shown in the left-hand and right-hand panel, respectively. For the coding of the velocity of the particles and the local density see figure~\ref{fig:small}. }
  	\label{fig:inter3}
  \end{figure}

 In the intermediate case, however, the chirality leads to a  synchronized motion that is completely different from the synchronized motion in the case of $\theta_{\chi}=0^\circ$ and $\theta_{\chi}=8^\circ$. Namely, in our system, a significant fraction   of the active objects  self-assembles into a vortex.
 Typical   configurations are shown in figure~\ref{fig:inter1} and~\ref{fig:inter3} for $\theta_\chi=1^\circ$ and $\theta_\chi=3^\circ$.
 For $\theta_\chi=2^\circ$, the configuration is similar to the two above cases, and the vortex takes on an intermediate size. In a system with $N$ and $L$ being  increased several times under the constraint $\rho=0.5$, a number of vortexes appears. 

\begin{figure}[h!]
	\centering
	\centering\includegraphics[scale=0.3]{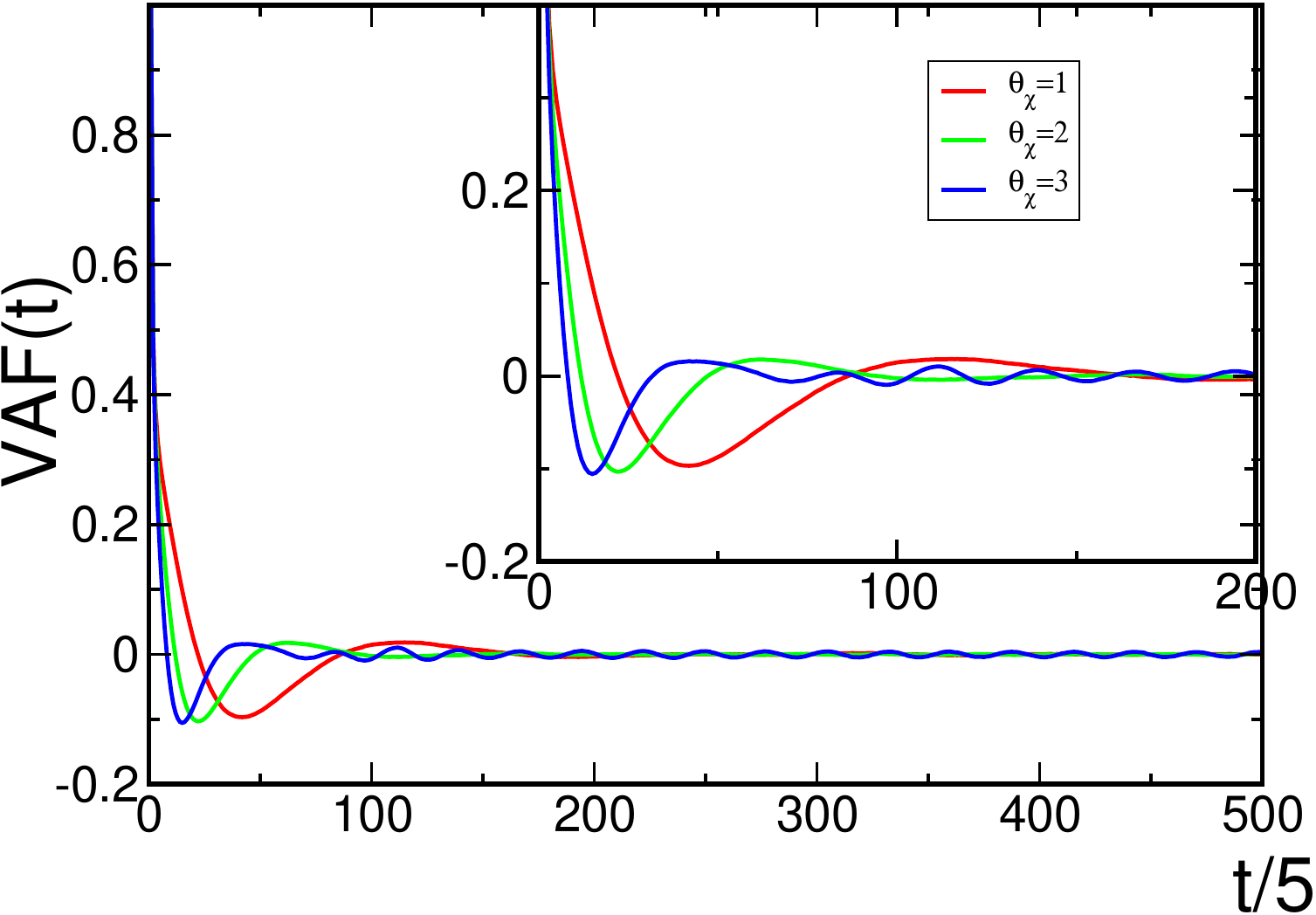} 
	\caption{(Colour online) The 
		velocity autocorrelation function [see (\ref{e:VAF})] in the system with the density $\rho=0.5$ and the number of particles  $N=810000$.  Red, green and blue lines correspond to the degree of chirality $\theta_\chi= 1.0^\circ, 2.0^\circ, 3.0^\circ$, respectively.  In the inset, ${\rm VAF}(t)$ is shown for short time of evolution, $t\leqslant 1000$. Time is divided by 5,  because we collected the data points every 5 time steps.   
	}\label{fig:VAF}  
\end{figure}

\begin{figure}[htpb]
	\centering
	\centering\includegraphics[scale=0.3]{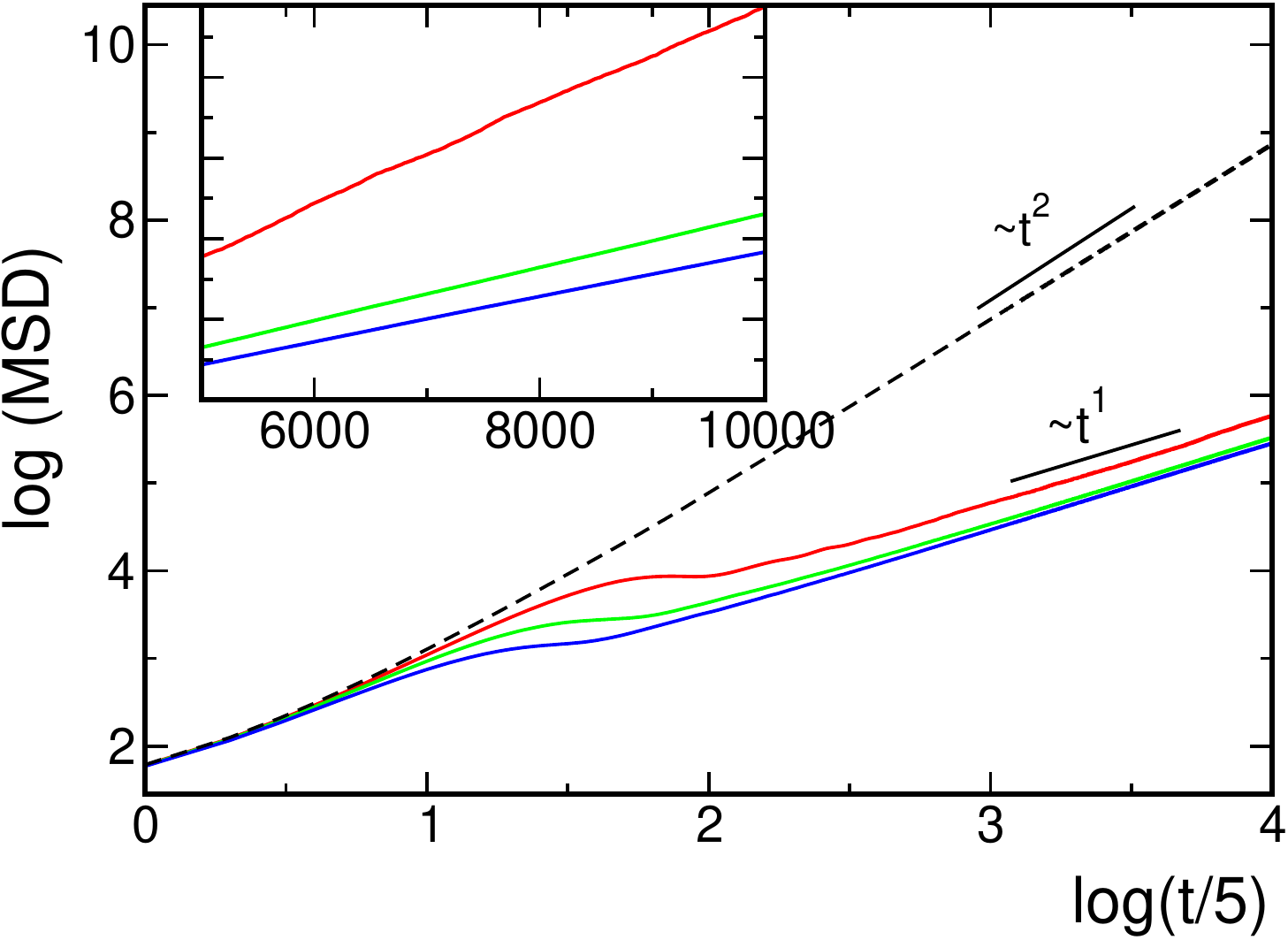}
	\caption{\label{fig:MSD}(Colour online) The MSD function. The density $\rho=0.5$ and the number of particles  $N=810000$. Red, green and blue lines correspond to the degree of chirality $\theta_\chi= 1.0^\circ, 2.0^\circ, 3.0^\circ$, respectively. The black dashed line corresponding to $\theta_{\chi}=0$ is included for comparison. In the inset $\langle \Delta r(t)^2\rangle$ is shown as a function of $t$ for $t>5500$. Time is divided by 5,  because we collected the data points every 5 time steps. }  
\end{figure}

  Formation of the vortexes is clearly seen in figure \ref{fig:inter1} and \ref{fig:inter3} by visual inspection. In order to analyse the evolution of the system with self-assembled vortexes on a quantitative level, we calculate the velocity autocorrelation function and the mean square displacement, equations~(\ref{e:MSD}) and (\ref{e:VAF}), for $\theta_{\chi}=1^\circ,2^\circ,3^\circ$, for all active objects, both inside and outside the vortex. 
  The results are shown in figure~\ref{fig:VAF} and \ref{fig:MSD} as functions of $t/5$, because we collected the data points every 5 time steps.
  
  Let us first note that in the absence of the noise ($\theta_{\rm max}=0^\circ$) an isolated object with the chirality~$\theta_{\chi}$  returns to the initial position after time $T=360 ^\circ/\theta_{\chi}=360, 180, 120$, and the radius of the circular motion is $r_f=T/(2\piup)\approx 57.3,28.6,19$ for $\theta_{\chi}=1^\circ,2^\circ,3^\circ$, respectively. The presence of different objects and the noise, however, have a significant effect on the time evolution. Interesting information can be obtained from the properties of the ${\rm VAF}(t)$ for relatively short times. As shown in figure~\ref{fig:VAF}, in each case the ${\rm VAF}(t)$ that is equal to properly averaged $\cos[\theta_i(t)-\theta_i(0)]$ [see (\ref{e:VAF})], is a smooth function with damped oscillations. The shape of the ${\rm VAF}(t)$ indicates that the vortex rotates as a whole with the period~$T$ that is approximately equal to the time between the first and the third zero of the ${\rm VAF}(t)$, or twice the time between the first minimum and the next maximum. We obtain $T\approx 276, 144, 110 $  for $\theta_{\chi}=1^\circ,2^\circ,3^\circ$, respectively. The smaller period of the circular motion in the vortex suggests that the trapped particles move on average on a circle with the radius $r<r_f$. Therefore, the angle $360^\circ$ is reached after shorter time than in the circular motion of the isolated particle without the noise. Indeed, the density in the vortex takes on  the maximum for $r\approx  36, 15.1, 9.46$  for $\theta_{\chi}=1^\circ,2^\circ,3^\circ$ that is smaller than the radius~$r_f$ of the circular motion of isolated particle without the noise. 
  
In figure~\ref{fig:VAF}, a rather fast decay of  ${\rm VAF}(t)$ can be observed, whereas the vortexes are stable for a long time, much larger than the decay time of ${\rm VAF}(t)$. This is because some objects are constantly absorbed in the vortex, and  some other objects constantly escape from its periphery. The objects escape because the density decreases with the distance from the center of the vortex, and at the periphery there are not enough particles to fix the orientation of a significant fraction of the objects.  The absolute value of the maximum is considerably smaller than the absolute value of the minimum of the ${\rm VAF}(t)$. It means that a considerable fraction of the objects forming the vortex escapes from it before returning to the initial orientation. At the same time, new objects enter the vortex.
  
As seen in figure~\ref{fig:VAF},  the period of rotation decreases considerably with increasing chirality, and the minimum becomes slightly deeper, consistent with decreasing size and increasing density of the vortex. Apart from that, the short-time evolution in each considered case is similar. For larger time, however, oscillations with a smaller period are present only for $\theta_{\chi}=3^\circ$, where the vortex takes on  the smallest size. This behavior may be due to the  presence of objects trapped in the vortex in the inner ring, where density is large enough to keep the circular movement of a considerable fraction of the objects. The period of this movement, $T_{in}=2\piup r_{in}/\bar v\approx 59$,  where $\bar v$ is the average speed taken over a few time steps  (figure~\ref{fig:VAF}) is smaller than $T\approx 110$, because the objects in the outer rings escape and do not contribute to the oscillatory behavior of  ${\rm VAF}(t)$ for such long times.

The properties of the long-time evolution of the active objects can be inferred from  the shape of the ${\rm MSD}(t)$ shown in figure~\ref{fig:MSD}. When $\theta_{\chi}=0^\circ$,  a ballistic evolution with $\langle \Delta r(t)^2\rangle\sim t^2$  is clearly seen for $\log(t/5)>2$, i.e., for $t>10^3$. For $\theta_{\chi}$ between $1^\circ$ and $3^\circ$, the  $\log [{\rm MSD}(t)]$ has a characteristic shape with three time regimes.  For short times, the slope is nearly constant and larger than 1. For intermediate times, $ t\sim 10^2$--$10^3$,   the slope takes on a very small value and increases to $1$ for large time of evolution, $t>10^3$.  The almost flat part of the $\log[{\rm MSD}(t)]$ moves to smaller time for increasing chirality.  Similar behavior of MSD has been observed in~\cite{sahala:25:0}, in the case of an inertial chiral active Ornstein--Uhlenbeck particle moving on a two-dimensional surface.

%

  $\langle \Delta r(t)^2\rangle\propto t$ for  $t>10^3$ indicates a diffusive behavior for $1^\circ\leqslant\theta_{\chi}\leqslant 3^\circ$.  
 The slopes of  $\langle \Delta r(t)^2\rangle$ for large $t$ on the linear scale (see the inset in figure~\ref{fig:MSD})  show differences between the diffusion constant $D$ for different chirality. From the relation  $\langle \Delta r(t)^2\rangle= 4D t$ for a 2D system and from the lines shown in the inset in figure~\ref{fig:MSD} we obtain 
  the results $D=3,1.65$ and $1.38$ for $\theta_{\chi}=1^\circ,2^\circ,3^\circ$, respectively.

  For the chirality $1^\circ\leqslant \theta_{\chi}\leqslant 3^\circ$, a significant fraction of the active objects  self-assembles into a long-lived vortex and the remaining objects form small clusters homogeneously distributed in the simulation box.
  The decrease of $D$ for increasing $\theta_{\chi}$ is consistent with a larger overall time spent by the objects in the vortex with a decresing size for increasing $\theta_{\chi}$, because the objects that  are temporarily trapped in the vortex move in the restricted space. 
  
  The average displacement  after the same time of evolution  decreases with increasing chirality. In the case of the chiral objects, the displacement is considerably smaller than in the case of the objects without chirality. 
  From figure~\ref{fig:MSD} we can see that for $t\sim 10^4$, we have $\sqrt{\rm MSD}\sim 3000$ and $\sqrt{\rm MSD}\sim  300$ for $\theta{\chi}=0^\circ$ and $\theta{\chi}=1^\circ$, respectively. Note that the noise in our model is between $-60 ^\circ$ and $60 ^\circ$, and the deviation from the symmetry of the noise as small as $1 ^\circ$ leads to quite different displacement of the objects after long time of evolution. 

 \section{Conclusions
 \label{conclusions}}
   
   We have found that the internal chirality of active objects considerably influences the synchronized motion that was observed in the Vicsek model  for a very large number of active objects moving on a surface. In our modification of the Viscek model, the chirality is modelled by a systematic turn to the right by the angle $\theta_{\chi}$ added to the  random deviation  from the  average local direction of motion that can occur  in the range from  $-\theta_{\rm max}$ to $+\theta_{\rm max}$ [see (\ref{theta_chi})]. In our simulations
    $\theta_{\rm max}= 60 ^\circ$, $N=810000$ and $\rho=0.5$.
   
   The bands travelling in the same direction in the case of the symmetrical noise (with $\theta_{\chi}=0^\circ$) are destroyed by very small chirality, as we have shown for the systematic turn to the right by the angle $\theta_{\chi}=0.5 ^\circ$, i.e.,  only for 
   $\theta_{\chi}/\theta_{\rm max}=1/120$. 
   The relatively large chirality, $\theta_{\chi}=8 ^\circ$, leads to the formation of approximately circular flocks of objects  that rotate as a whole, and all members of the flock move approximately in the same direction.
   
   The  vortexes appear and remain stable in the stationary state for  small chirality, as we have shown for  $\theta_{\chi}=1^\circ,2^\circ,3^\circ$ that is between $\theta_{\chi}/\theta_{\rm max}=1/60$ and $\theta_{\chi}/\theta_{\rm max}=1/20$. The shape of VAF as well as visual inspection of the time evolution show that the stability of the vortex follows from the compensation of the constant escape of the objects from the vortex by the constant inflow of other objects. Trapping of the objects in the vortexes drastically modifies the long-time evolution that changes from the ballistic one in the case of bands travelling in the same direction for $\theta_{\chi}=0^\circ$ to the diffusive one, with the diffusion constant decreasing with increasing chirality. 
 
 Let us finally mention that   similar results were obtained in diffrent models of active matter~\cite{liebchen:17:0,kruk:20:0,ventejou:21:0}. This suggests universality of the vortex fomation by self-propelled objects with small chirality, where by the universality we  mean independence of this phenomenon  of the details that are diferent in different models. 
  In this context it is worth  
 	mentioning that depending on  possible applications, the internal chirality can be an advantage or a disadvantege. 
 	If a movement in the same direction of  active objects such as microrobots or drones that adjust the velocity to the average direction of close neighbors is required, then even small asymmetry in their construction should be avoided. On the other hand, vortex formation can find various applications. For example, when the active objects adsorb impurities in the system, they become localized in the vortex where the density takes a maximum, and can be easily removed. Also, as observed in bacterial colonies, vortex formation can generate enough torque to rotate a rigid rod, by which micromechanical actuation can take place. 
 	Our results show for which degree of chirality the formation of a vortex can be expected, and can guide a design of biomimetic active objects with synchronised motion on demand. 
 	
 	In future studies it is important to investigate the role of boundary conditions and obstacles that are commonly present in nature and the role of the number  as well as different chiralities of active objects. Practical open questions concern in particular the role of chirality in (i) transporting microscopic cargo along specific paths, (ii)
 	directed flows in microchannels, (iii) controlling the mixing or segregation of different components in a fluid and (iv)  designing autonomous transport systems at the microscale.

\section*{Acknowledgement}

This article is dedicated to the memory of prof. Stefan Soko\l owski, our colleage and supervisor of the master thesis of WTG. 
 The article by Gozdz, Patrykiejew and Sokolowski~\cite{gozdz:90:0}, initiated the scientific career of WTG.

\bibliographystyle{cmpj}
\bibliography{ref-vortex(2)}


\newpage
\ukrainianpart

\title{Формування вихорів у моделі Вічека з внутрішньою хіральністю саморухомих об'єктів}
%
%
\author{В. T. Гоздзь, A. Цях }

\address{Інститут фізичної хімії Польської академії наук, 01-224 Варшава, Польща}

\makeukrtitle

\begin{abstract}
	\tolerance=3000%
	Вплив внутрішньої хіральності на колективний рух великої кількості активних об'єктів вивчається за допомогою моделювання в рамках  модифікованої моделі Вічека. Ми додаємо фіксований кут до шуму та розглядаємо малі відношення $p$ цього кута до максимального відхилення від середнього локального напрямку руху. Коли вищезгадане відношення становить $p=1/120$, рухомі зони, що спостерігаються із симетричним шумом, руйнуються, і з'являються малі зони, що рухаються в різних напрямках. Кругові обертові скупчення об'єктів з однаковою орієнтацією формуються при $p=1/7.5$. Стабільні вихори у стаціонарному стані були виявлені від $p=1/60$ до $p=1/20$. Автокореляційна функція швидкостей показує рівновагу між припливом до та відпливом з вихору. Тимчасове захоплення об'єктів у вихорі значною мірою впливає на довгочасову еволюцію системи. Балістична поведінка для симетричного шуму змінюється на дифузійну поведінку для хіральності, що призводить до виникнення вихорів.
	\keywords саморухомі об'єкти, внутрішня хіральність, колективний рух, модель Вічека
	
\end{abstract}

\lastpage
\end{document}